\documentclass{article}

     \usepackage[nonatbib, final]{neurips_2019}

\usepackage[utf8]{inputenc} 
\usepackage[T1]{fontenc}    
\usepackage{hyperref}       
\usepackage{url}            
\usepackage{booktabs}       
\usepackage{amsfonts}       
\usepackage{nicefrac}       
\usepackage{microtype}      
\usepackage{float}
\usepackage{graphicx}

\usepackage{cite}

\title{Portable system for the prediction of anemia based on the ocular conjunctiva using Artificial Intelligence}

\author{
  \textbf{Bryan Saldivar-Espinoza$^1$, Dennis Núñez-Fernández$^1$, Franklin Porras-Barrientos$^1$} \\ \textbf{Alicia Alva-Mantari$^1$, Lisa Suzanne Leslie$^2$, \iffalse Nicolas G. Rey De Castro$^3$, \fi Mirko Zimic$^1$}
  \\
  $^1$Laboratorio de Bioinformática y Biología Molecular, Universidad Peruana Cayetano Heredia, Peru \\ 
  $^2$Weill Cornell Medicine, Cornell University, USA \\ 
  \texttt{\{bryan.saldivar.e, dennis.nunez, franklin.barrientos.p\}@upch.pe} 
  \\
  \texttt{alicia.alva.m@upch.pe, lsl2001@med.cornell.edu, \iffalse nicolasreydecastro@doctors.org.uk, \fi mirko.zimic@upch.pe} 
}

\begin{document}

\maketitle

\begin{abstract}
Anemia is a major health burden worldwide. Examining the hemoglobin level of blood is an important way to achieve the diagnosis of anemia, but it requires blood drawing and a blood test. In this work we propose a non-invasive, fast, and cost-effective screening test for iron-deficiency anemia in Peruvian young children. Our initial results show promising evidence for detecting conjunctival pallor anemia and Artificial Intelligence techniques with photos taken with a popular smartphone.
\end{abstract}

\section{Introduction}

Anemia is a global public health problem with major consequences for human health. It affects 24.8\% of the world's population \cite{1} and 43,5\% of Peruvian children \cite{2}. It is assessed by the level of hemoglobin (Hb), which is the most reliable indicator. Current clinical methods primarily rely on blood extraction, which demands high labor, instrumentation costs, and is a time-consuming procedure which exposes the personal to blood-transmissible diseases.

Over the last decades, various algorithms and devices have been designed in order to determine the state of anemia based on the color of the conjunctiva. In \cite{3, 4}, a good correlation between the level of Hb and the conjunctival pallor was obtained via a smartphone-based system. In \cite{5}, the authors analyzed color features of the palpebral conjunctiva based on a standard gray card with the aim of correcting the colors of the images. In their experiment, they found a moderate correlation between the hub and the color features. In \cite{6}, the authors taken into consideration the spectral reflectance of the conjunctiva for clear anemia cases, but the equipment used are not commonly available and highly expensive. In \cite{7}, it was proposed a comparison two different classification methods based on a support vector machine and a deep neural network, obtaining relatively good performance. 

In this work we propose a computer-based method that can predict anemia by just analyzing images of palpebral conjunctiva. Therefore, this method is fast, inexpensive, globally applicable, and popular enough to replace a physician's visual examination for the diagnosis of anemia. Such a system would be very useful in rural areas of developing countries, where medical resources are limited.

\section{Methodology}

\begin{figure}[H]
  \centering{\includegraphics[width=120mm]{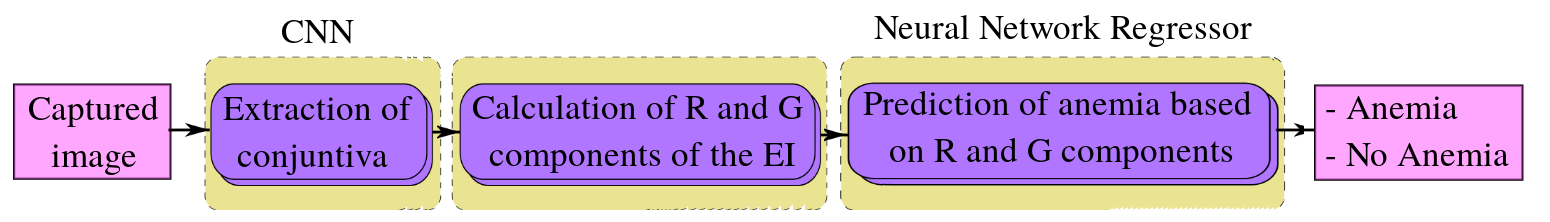}}
  \caption{Diagram for the proposed system}
  \label{diagram}
\end{figure}

The proposed system works with images captured from a smartphone camera. The images are processed on UPCH servers to predict anemia and the result is returned to the smartphone. Conjunctiva extraction is performed via automatic segmentation using CNNs, then R and G components of the Erythema Index (EI) are calculated. Finally, using these values and a neural network regressor, the prediction of Anemia is estimated. Fig.~\ref{diagram} shows the diagram of the system.

The dataset was collected in three state medical centers from Lima with its respective authorization, obtaining about 300 initial samples from young children. In order to have a gold standard value, the clinical method was used to label each sample with its Hb value. Conjunctiva extraction was done by using a CNN with 35 layers. For conjunctiva segmentation we used a 7 layer CNN with the segmented conjunctiva as the target (3 channels). Then, for conversion to mask, a histogram of 128 values was calculated. Based on the conjunctiva region, the EI was determined using the equation reported by Yamamoto et al \cite{8}: $ EI = log(R) - log(G) $ where R and G are the brightness of the conjunctiva in red a green channel. In order to normalize the images due ambiental lighting, each channel’s brightness were adjusted by multiplying its brightness by 200/MB where MB is the mean brightness of the color calibration card’s white square. Finally, based on the R and G components, the Hb value is predicted by a neural network regressor, and classified as anemia or no anemia by employing the predicted Hb value. We use Python and Tensorflow for the implementation.

\section{First Results}

\begin{figure}[H]
  \centering{\includegraphics[width=90mm]{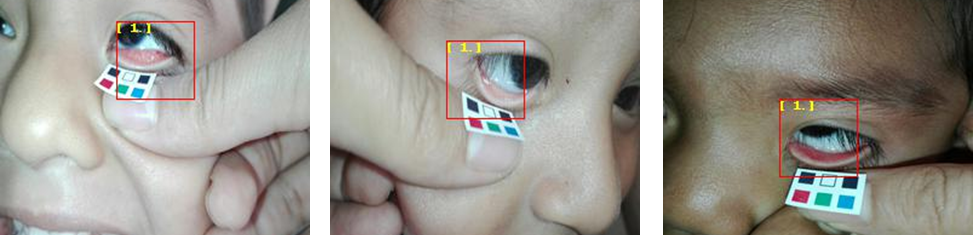}}
  \caption{First results of the automatic conjunctiva detector}
  \label{segmentation}
\end{figure}

The improvement of the system is under development and the final results no prepared, however the initial results are presented in this work and show promising evidence to detect anemia. As explained in the methodology section, the first part of the system is the extraction of the conjunctiva by CNNs. As we can see in the Fig.~\ref{segmentation}, the implemented CNN achieves successful results.

Furthermore, taking into consideration different cut-off points for anemia prediction, the initial proposed neural network regressor achieves promising results. As the Table.~\ref{hb_table} shows, accuracy for two classes, sensibility and specificity are evaluated for each cut-off point. The best results are obtained from the cut-off point of Hb = 11, which gives a sensitivity of 77.58\%.

\begin{table}[H]
\small
\caption{Results of the neural network regressor for different cut-off points [\%]}
\label{hb_table}
\centering
\begin{tabular}{llll}
\toprule
Cut-off point  & Hb = 9 & Hb = 10 & Hb = 11 \\
\midrule
Accuracy & 93.53 & 77.07 & 42.96 \\
Sensitivity & 22.49 & 52.21 & 77.58 \\
Specificity & 94.20 & 78.40 & 36.03 \\
\bottomrule
\end{tabular}
\end{table}

\section{Conclusions}

In this work we demonstrated that our system is capable to predict anemia with a sensitivity of 77.58\% via images captured from a popular smartphone and using Artificial Intelligence techniques. Although the project is still under development, more samples are being collected and some changes are still being made, the initial evidence show promising results. In this way, our system provides a fast, cost-effective, and globally useful screening tool for anemia evaluation, especially in the Peruvian rural areas where medical resources are limited.



\clearpage

\small

\bibliography{sample}{}

\begin{thebibliography}{1}

\bibitem{7}
Yi-Ming Chen, Shaou-Gang Miaou, and Hongyu Bian.
\newblock Examining palpebral conjunctiva for anemia assessment with image
  processing methods.
\newblock {\em Computer Methods and Programs in Biomedicine}, 137:125 -- 135,
  2016.

\bibitem{2}
Instituto~Nacional de~Estadística~e Informática.
\newblock Encuesta demográfica y de salud familiar 2016 nacional y regional
  (endes 2016).
\newblock 2017.

\bibitem{4}
G.~{Dimauro}, L.~{Baldari}, D.~{Caivano}, G.~{Colucci}, and F.~{Girardi}.
\newblock Automatic segmentation of relevant sections of the conjunctiva for
  non-invasive anemia detection.
\newblock In {\em 2018 3rd International Conference on Smart and Sustainable
  Technologies (SpliTech)}, pages 1--5, June 2018.

\bibitem{3}
G.~{Dimauro}, D.~{Caivano}, and F.~{Girardi}.
\newblock A new method and a non-invasive device to estimate anemia based on
  digital images of the conjunctiva.
\newblock {\em IEEE Access}, 6:46968--46975, 2018.

\bibitem{6}
Oleg Kim, John McMurdy, Gregory Jay, Collin Lines, Gregory Crawford, and Mark
  Alber.
\newblock Combined reflectance spectroscopy and stochastic modeling approach
  for noninvasive hemoglobin determination via palpebral conjunctiva.
\newblock {\em Physiological Reports}, 2(1):e00192, 2014.

\bibitem{1}
World~Health Organization et~al.
\newblock Worldwide prevalence of anaemia 1993-2005: Who global database on
  anaemia.
\newblock 2008.

\bibitem{5}
Selim Suner, Gregory Crawford, John McMurdy, and Gregory Jay.
\newblock Non-invasive determination of hemoglobin by digital photography of
  palpebral conjunctiva.
\newblock {\em Journal of Emergency Medicine}, 33(2):105--111, Aug 2007.

\bibitem{8}
Tadamasa Yamamoto, Hirotsugu Takiwaki, Seiji Arase, and Hiroshi Ohshima.
\newblock Derivation and clinical application of special imaging by means of
  digital cameras and image j freeware for quantification of erythema and
  pigmentation.
\newblock {\em Skin Research and Technology}, 14(1):26--34, 2008.

\end{thebibliography}
\bibliographystyle{plain}

\end{document}